\documentclass[conference]{IEEEtran}
\IEEEoverridecommandlockouts

\usepackage{cite}
\usepackage{amsmath,amssymb,amsfonts}
\usepackage{algorithmic}
\usepackage{graphicx}
\usepackage{tikz}
\usetikzlibrary{positioning,arrows.meta}
\usepackage{textcomp}
\usepackage{xcolor}
\usepackage{tabularx}
\usepackage{booktabs} 
\usepackage{multirow} 
\usepackage{pgfplots}
\usepackage{subcaption}
\usepackage{siunitx}

\def\BibTeX{{\rm B\kern-.05em{\sc i\kern-.025em b}\kern-.08em
    T\kern-.1667em\lower.7ex\hbox{E}\kern-.125emX}}
\begin{document}

\title{Multi-Agent Reinforcement Learning for C-V2X RAT Selection \\
\thanks{}
}

\author{\IEEEauthorblockN{Moritz Schaffenroth, Uwe Kölbel, Heike Lepke, Alexander Prinz, Alfred Höß}
\IEEEauthorblockA{\textit{Department of Electrical
Engineering, Media and Computer Science} \\
\textit{Ostbayerische Technische Hochschule Amberg-Weiden}\\
Amberg, Germany \\
\{m.schaffenroth, u.koelbel, h.lepke, a.prinz, a.hoess\}@oth-aw.de}
}

\maketitle

\begin{abstract}
    Vehicles are increasingly equipped with advanced V2X communication capabilities.
    While early V2X apps utilized services such as Cooperative Awareness Messages, recent developments have allowed more advanced applications including cooperative driving, shared perception, and sensor-sharing services.
    The broader mix of applications leads to heterogeneous requirements for latency and reliability. At the same time multiple communication technologies for V2X are available with pros and cons.
    Hybrid V2X communication can exploit the distinct advantages at the right moment to fulfill the requirements of the applications. This work studies the decision problem between cellular Uu link, NR-V2X PC5 sidelink, and the simultaneous use of both channels.
    We address this problem by using the multi-agent reinforcement learning algorithm MAPPO and compare it to five baselines consisting of a deep reinforcement learning (DRL) approach, a static decision tree approach and static channel selection strategies.
    The methods are evaluated in an urban scenario and with a set of selected communication use cases.
    The evaluation results show that when compared to the DRL approach, the on-time delivery ratio improves from 0.508 to 0.535 in a single-controlled-vehicle setting and from 0.548 to 0.567 when all vehicles follow the learned policy and reduces the training time by half.
    The gains result mainly from the advanced applications scenarios, as opposed to scenarios involving exclusively CAM messaging. This indicates future applications will benefit from such adaptive communication strategies and that multi-agent modelling is useful for addressing the underlying decision problem.

\end{abstract}


\section{Introduction}
The landscape of V2X enabled applications is rapidly expanding. Future use cases like see-through for passing and cooperative perception have heterogeneous quality-of-service (QoS) requirements and communication with a single technology e.g., PC5 or Uu link can lead to performance limitations in terms of latency, reliability, or coverage.

While PC5 enables direct low latency communication with ultra-low latency: $\sim$\qty{1} milliseconds \cite{s21030843}, the Uu link (5G) provides advantages such as higher bandwidth, extended communication range, connectivity to remote infrastructure but demands the availability of cellular network coverage.
Recent work by Khalid et al.\cite{fi16040107} has shown that hybrid V2X selection can improve the one-way end-to-end latency, reliability and packet reception ratio of the communication when compared to static technology selection. They proposed a selection algorithm based on a decision tree to select between 5G/UU, C-V2X, and dedicated short range communication (DSRC) based on CAM-message statistics. This approach is explainable but might be limited under non-stationary situations as it occurs in multi-agent settings. Yacheur et al.\cite{10199400} formulate the decision problem as deep reinforcement learning (DRL) task and compare their solution to single radio access technology (RAT) selection and a multiple-criteria decision analysis (MCDM) algorithm and assess it under two situations: Congested and non-congested. They find that in their scenario the packet reception rate (PRR) increases from \qty{66}{\percent} to more than \qty{90}{\percent}. This work highlights the advantage use of DRL for the RAT selection problem.

Mancini et al. \cite{10881343} study joint access selection with DSRC and visible light communication (VLC) and a complex state that e.g., considers distances between vehicles. They train vehicles under three different situations: urban, countryside, and highway and combine the models via Federated Deep Reinforcement Learning (FedDRL). Their observation was a more stable training process and generalization over different scenarios.

However, the hybrid Uu/PC5 selection is a multi-agent setting where multiple agents compete for limited resources such as channel occupancy time and influence each other through their channel choice.
This competition is becoming increasingly relevant as the number of connected vehicles and number of applications grows.

Prior work\cite{9868865} on a related field has applied multi agent reinforcement learning to optimize radio resource selection focusing on mitigating packet collisions and improving spectrum efficiency. However, their approach focuses on a single technology whereas we focus on the selection between multiple technologies such as Uu link and PC5.

This paper targets the channel selection between Uu and PC5 in an urban ns-3/SUMO simulation. In particular we compare a centralized RL approach and a mixed cooperative-competitive environment multi-agent reinforcement learning (MARL) formulation.

The main contributions are as follows:

\begin{itemize}
\item a MARL description of the problem of hybrid Uu/PC5 mode selection using a simulation stack based on the VaN3Twin\cite{pegurri2026van3twinmultitechnologyv2xdigital} ecosystem and observations and a reward function based on QoS requirements that penalizes hybrid duplication
\item The MARL formulation is evaluated using a multi-agent proximal policy optimisation (MAPPO) model\cite{10.5555/3600270.3602057} in a  ns3/SUMO digital-twin simulation environment with a heterogeneous application mix and a varying number of vehicles. This model achieved a higher mean on-time delivery ratio than a centralised double deep Q-network (DDQN) baseline inspired by \cite{10199400}.
\end{itemize}

\section{Simulation Stack}
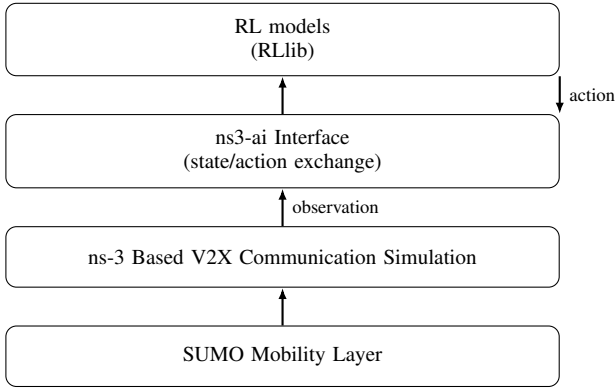
\begin{figure}[t]
\centering
\begin{tikzpicture}[
  node distance=5mm,
  box/.style={
    draw,
    rounded corners,
    align=center,
    text width=0.8\columnwidth,
    minimum height=8mm,
    inner ysep=2mm,
    font=\footnotesize
  },
  arr/.style={-{Latex[length=1.5mm]}, thick}
]
  \node[box] (py) {RL models\\(RLlib)};
  \node[box, below=of py] (bridge) {ns3-ai Interface\\(state/action exchange)};
  \node[box, below=of bridge] (ns3) {ns-3 Based V2X Communication Simulation};
  \node[box, below=of ns3] (sumo) {SUMO Mobility Layer};

  \draw[arr] (sumo) -- (ns3);
  \draw[arr] (ns3) -- node[right, font=\scriptsize]{observation} (bridge);
  \draw[arr] (bridge) -- (py);
  \draw[arr] (py.south east) to[out=-90,in=90] node[right, font=\scriptsize]{action} (bridge.north east);
\end{tikzpicture}
\caption{Simulation stack.}
\label{fig:co-sim-stack}
\end{figure}

The digital twin for the communication simulation is built upon the ms-van3t / VaN3Twin ecosystem\cite{RAVIGLIONE202470}\cite{pegurri2026van3twinmultitechnologyv2xdigital}. This framework integrates a microscopic mobility simulation using the traffic simulator SUMO \cite{dlr71460}. The communication simulation is based on the ns-3 network simulator\cite{ns3}. On top of that stack we implemented our own custom channel selection model. In our setup, we used ns3-ai\cite{10.1145/3389400.3389404} to couple the C++ based ns-3 simulation with Python based reinforcement learning to exchange observations and actions. To support large scale parallel execution, we modified ns3-ai to support multiple simulation instances. The top layer of the simulation stack consists of rllib \cite{pmlr-v80-liang18b} based code which performs both inference and training for the channel selection models. An overview of the complete stack is provided in Fig. \ref{fig:co-sim-stack}.

\section{RL formulation}
We consider two potential RATs for this work. The observations are a 10-dimensional vector containing Uu and PC5 channel busy ratios, Uu SINR, ego speed, and six application-level parameters: priority, latency target, message size, periodic flag, interval, and communication range.\\
The discrete action space consists of three options and is given as follows:

\[
A = \begin{cases}
0, & \text{Uu link} \\
1, & \text{PC5} \\
2, & \text{Uu link \& PC5}
\end{cases}
\]


Our implementation follows the Centralized Training with Decentralized Execution (CTDE) paradigm, where during training a critic has access to global state information (global on-time delivery ratio, number of vehicles in range, global number of vehicles, channel busy rate, packet reception ratio per channel) and the combined reward of all agents. We use a shared actor and assume all vehicles face the same decision problem. This architecture has the following advantages: Improved sample efficiency and better generalization.
Every vehicle decides for every message. Since this occurs for each vehicle at different times, the execution is turn-based.

We compare our approach to a DDQN model trained with a different reward function proposed by \cite{10199400}. They consider in their reward message reception acknowledgments as well as a performance satisfaction measurement, and link quality, where the latter is a comparison of current and previous signal-to-noise-plus-interference ratio (SNIR) values. Since the paper does not clearly state in which setting the model was trained and to have a comparable setting we trained it as a shared policy and in a non-stationary setting. The observation space is identical for both models.

We define a simpler reward which is a  composite function that balances individual QoS with overall efficiency. The reward consists of a local component that reflects the agent's self-interest and a global component that incentivizes cooperation:
\begin{equation}
r_t^i =
(1-\lambda)\left(
\sum_{m \in \mathcal{R}_t^i} w_m \frac{s_m}{n_m}
-
\gamma_{\mathrm{dup}} h_t^i
\right)
+
\lambda d_t ,
\qquad \lambda \in [0,1],
\end{equation}

\noindent where\\

\begin{tabularx}{\columnwidth}{@{}l@{\quad}X@{}}
$\mathcal{R}_t^i$ & set of messages generated by agent $i$ whose outcome is resolved at timestep $t$, \\
$w_m$ & priority weight of message $m$, \\
$s_m$ & number of intended receivers that obtained message $m$ within its latency deadline, \\
$n_m$ & number of intended receivers within the application's communication range, \\
$h_t^i$ & number of hybrid duplicate receptions caused by agent $i$ at timestep $t$, \\
$\gamma_{\mathrm{dup}}$ & penalty coefficient for hybrid duplication, \\
$d_t$ & global on-time delivery ratio across all controlled agents at timestep $t$. \\
\end{tabularx}\\

The weighting between global and local reward $\lambda$ is $0.3$ in our experiments. Hybrid duplicates refer to packets received via both channels, as there could be also duplicates over the same channel in case of re-transmissions.

\section{Experimental Setup}
\subsection{Simulated scenario}
\hfil
\begin{figure*}[t]
\centering
\subfloat[Urban Map A]{
  \includegraphics[width=0.31\textwidth]{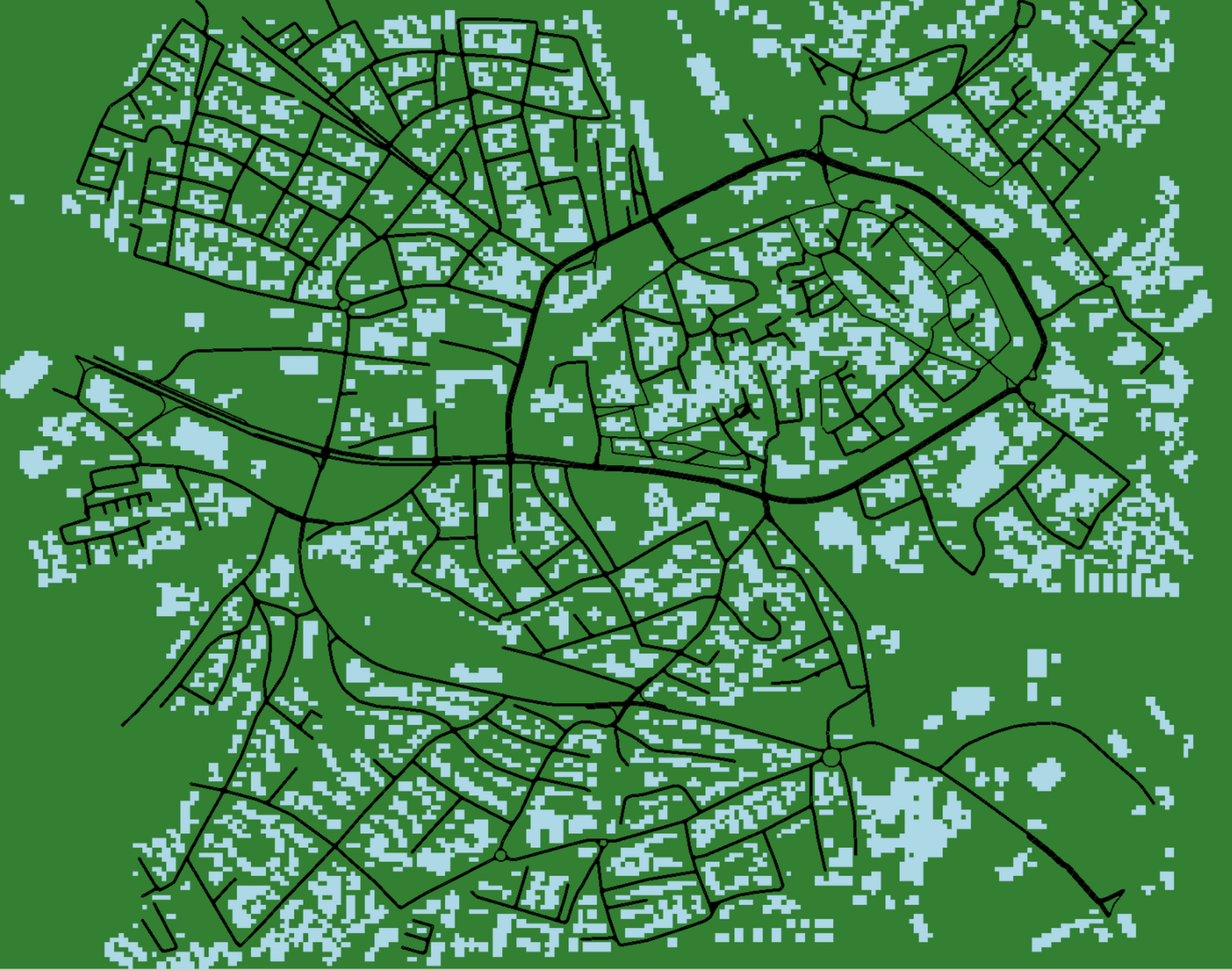}
}
\hfil
\subfloat[Urban Map B]{
  \includegraphics[width=0.31\textwidth]{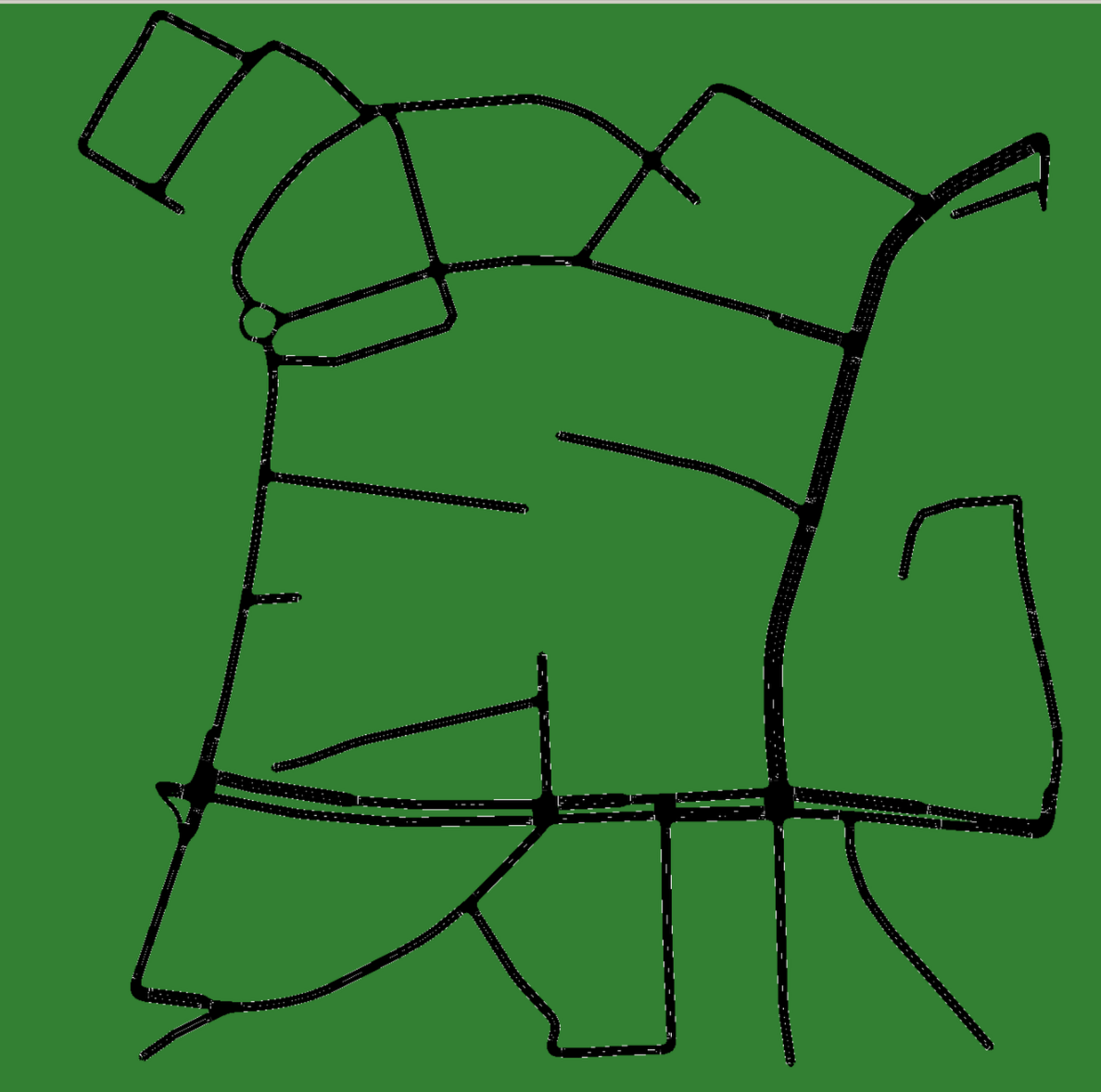}
}
\hfil
\caption{Road-network maps used in this study.}
\label{fig:sumo-maps}
\end{figure*}

During the training of the two reinforcement learning models, the Yacheur et al. DDQN model and our MAPPO model we use two SUMO road networks which we refer to as Urban Map A and Urban Map B. Random traffic demand is generated for both maps using four application-mix profiles and two traffic density factors. The traffic generation script samples vehicle routes and departure times stochastically, using insertion periods of 2.5, 3.5, 5, 8, and 10 seconds. The traffic demand factors scale those numbers. In both training and evaluation, the traffic simulation is first advanced to a defined start time so that vehicles are already present in the scenario before communication decisions are executed. Evaluation is performed on both maps and they are illustrated in Fig. \ref{fig:sumo-maps}. For every combination of scenario, app-mix and start time, the evaluation is run 10 times under identical random seeds for all methods. We evaluate the learned policies under two different levels of control: Single-controlled-vehicle, only one ego vehicle is controlled by the reinforcement learning algorithm and all other vehicles are non-adaptive traffic and All-vehicles-controlled, where every vehicle in the simulation runs the same policy. Here agents influence each other through channel occupancy, interference and hybrid duplication. In this scenario the real multi-agent nature of the problem shows. The app-mixes are as follows: CAM-only, Mixed, Mostly-Uu and Mostly-PC5. CAM-only is a scenario where only CAM messages are exchanged. The remaining app-mixes consist of nine selected V2X use cases from \cite{gaa2025cv2xvol1}, but have different default channel for non-adaptive traffic per app which leads to traffic that has different utilization on channels.
We consider the following evaluation settings, which lead to 16 evaluation configurations:
four app-mixes, two maps, two start times.

\subsection{Communication setup}

The compared methods use the same PHY/MAC settings. Uu at \qty{3.5}{\GHz} high frequency-band and \qty{80}{\MHz} bandwidth, and PC5 at \qty{5.89}{\GHz} high frequency-band and \qty{20}{\MHz} baseband using the NR-V2X Mode 2 and fixed modulation and coding scheme (MCS) of 10.

\subsection{Method comparison}

In our evaluation we compare the following methods:
\begin{itemize}
    \item static Uu-only RAT selection;
    \item static PC5-only RAT selection;
    \item static Hybrid RAT selection;
    \item a rule-based decision-tree baseline inspired by \cite{fi16040107};
    \item a centralized learned reference policy (DDQN) trained under the assumption of non-adaptive surrounding traffic;
    \item our MAPPO approach trained, where all vehicles simultaneously learn to coordinate;
\end{itemize}

We adjusted the rule-base decision-tree to map the DSRC channel to PC5 as we only consider PC5 as V2V communication channel in this work.
The DDQN model uses our observation space, but the Yacheur et al. reward signal as we use 5G-NR-V2X Mode 2 instead of DSRC.

\subsection{Training}

The training budget for both RL approaches is configured to be comparable. The MAPPO approach is trained on 60 iterations with 100-step episodes, which leads in our configuration to a total of $4.8 \times 10^{5}$ environment steps per training run.
The DDQN model uses also 60 iterations, but with a message receive target of 100, and the episode steps were capped at 200 steps which lead to a total of $\sim 6.7 \times 10^{5}$ environment-steps.
The DDQN checkpoints took on our local setup on average about 2.2 days to train. The MAPPO checkpoints in mean about 17.6 hours.

\subsection{Metrics}

\begin{table}[t]

\renewcommand{\arraystretch}{1.2}
\centering
\caption{Key performance indicators used for evaluation.}
\label{tab:kpi-definitions}
\small
\begin{tabularx}{\columnwidth}{@{}llX@{}}
\toprule
\textbf{KPI} & \textbf{Unit} &\textbf{Definition} \\
\midrule
On-time delivery & ratio & latency-compliant deliveries / expected packets \\
On-time reliability & ratio & latency-compliant deliveries / arrived packets \\
Mean delay & [s] & mean end-to-end delay over arrived packets \\
Hybrid duplicate rate & [\%] & share of receptions where the same message arrives via both Uu and PC5 \\
Best overall configurations & \# & grouped evaluation configurations in which a method has the highest mean on-time delivery ratio \\
\bottomrule
\end{tabularx}
\end{table}

The primary metric is \textit{on-time delivery ratio}, defined as deadline compliant deliveries divided by the expected packets. We also report \textit{on-time reliability}, \textit{mean delay of arrived packets} and \textit{hybrid duplicate ratio} as can be seen in Table \ref{tab:kpi-definitions}.

\section{Results}
  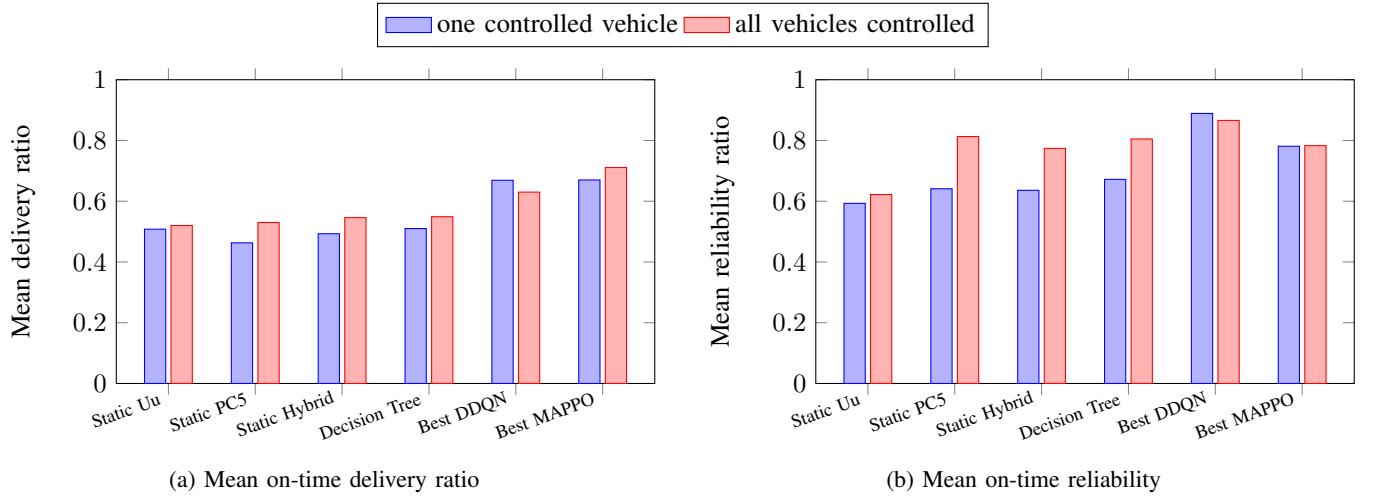
\begin{figure*}[t]
  \centering
  \begin{tikzpicture}
  \node at (0,0) {};
  \end{tikzpicture}
  \pgfplotslegendfromname{mainresultslegend}
  \vspace{2mm}

  \subfloat[Mean on-time delivery ratio\label{fig:main-delivery}]{
  \begin{tikzpicture}
  \begin{axis}[
  width=0.48\textwidth,
  height=5.6cm,
  ybar,
  bar width=8pt,
  ymin=0,
  ymax=1,
  symbolic x coords={Static Uu,Static PC5,Static Hybrid,Decision Tree,Best DDQN,Best MAPPO},
  xtick=data,
  xticklabel style={rotate=20,anchor=east,font=\scriptsize},
  ylabel={Mean delivery ratio},
  legend to name={mainresultslegend},
  legend columns=2,
  enlarge x limits=0.12,
  area legend
  ]
  \addplot coordinates {(Static Uu,0.508) (Static PC5,0.463) (Static Hybrid,0.493) (Decision Tree,0.510) (Best DDQN,0.669) (Best MAPPO,0.670)};
  \addplot coordinates {(Static Uu,0.520) (Static PC5,0.530) (Static Hybrid,0.546) (Decision Tree,0.549) (Best DDQN,0.630) (Best MAPPO,0.711)};
  \legend{one controlled vehicle,all vehicles controlled}
  \end{axis}
  \end{tikzpicture}
  }
  \hfill
  \subfloat[Mean on-time reliability\label{fig:main-reliability}]{
  \begin{tikzpicture}
  \begin{axis}[
  width=0.48\textwidth,
  height=5.6cm,
  ybar,
  bar width=8pt,
  ymin=0,
  ymax=1.0,
  symbolic x coords={Static Uu,Static PC5,Static Hybrid,Decision Tree,Best DDQN,Best MAPPO},
  xtick=data,
  xticklabel style={rotate=20,anchor=east,font=\scriptsize},
  ylabel={Mean reliability ratio},
  enlarge x limits=0.12
  ]
  \addplot coordinates {(Static Uu,0.593) (Static PC5,0.641) (Static Hybrid,0.636) (Decision Tree,0.672) (Best DDQN,0.889) (Best MAPPO,0.781)};
  \addplot coordinates {(Static Uu,0.622) (Static PC5,0.813) (Static Hybrid,0.774) (Decision Tree,0.805) (Best DDQN,0.866) (Best MAPPO,0.783)};
  \end{axis}
  \end{tikzpicture}
  }
  \caption{Means for on-time delivery ratio and on-time reliability. The baseline methods and the best DDQN/MAPPO model seeds are compared over 16 evaluation configurations per controlled-vehicle setting.}
  \end{figure*}

\begin{table}[t]
  \renewcommand{\arraystretch}{1.1}
  \centering
  \caption{Direct comparison between MAPPO and DDQN based on 10 models trained with different seeds each. Evaluation summary over 16 evaluation configurations per controlled-vehicle setting. Means are aggregated across configurations.}
  \label{tab:main-results}
  \scriptsize
  \begin{tabularx}{\columnwidth}{p{1cm}lp{1cm}p{.9cm}p{.5cm}p{1.3cm}}
  \toprule
  \textbf{Controlled vehicles} & \textbf{Method} & \textbf{Best overall configurations} & \textbf{Mean delivery ratio} & \textbf{Delay [ms]} & \textbf{Hybrid duplication [\%]} \\
  \midrule
  \multirow{2}{*}{one}
  & DDQN & \phantom{1}7 & 0.508 & \phantom{0}96 & 12.119 \\
  & MAPPO & \phantom{1}9 & 0.535 & 102 & \phantom{0}9.645 \\
  \midrule
  \multirow{2}{*}{all}
  & DDQN & \phantom{1}3 & 0.548 & \phantom{0}57 & \phantom{0}8.762 \\
  & MAPPO & 13 & 0.567 & \phantom{0}58 & \phantom{0}8.892 \\
  \bottomrule
  \end{tabularx}
  \end{table}

 \begin{table}[t]
  \caption{Winner counts by application mix and controlled vehicles. Winners are selected by the highest mean on-time delivery ratio.}
  \label{tab:method-win-counts}
  \centering
  \scriptsize
  \begin{tabularx}{\columnwidth}{p{1cm}lXXXX}
  \toprule
  \textbf{Controlled vehicles} & \textbf{App-mix} & \textbf{Static Uu-only} & \textbf{Static PC5-only} & \textbf{DDQN} & \textbf{MAPPO} \\
  \midrule
  \multirow{4}{*}{one}
  & CAM-only & 2 & 2 & - & - \\
  & Mixed & 1 & - & 1 & 2 \\
  & Mostly-PC5 & 1 & - & - & 3 \\
  & Mostly-Uu & - & - & 3 & 1 \\
  \midrule
  \multirow{4}{*}{all}
  & CAM-only & 3 & 1 & - & - \\
  & Mixed & - & - & 1 & 3 \\
  & Mostly-PC5 & - & - & 1 & 3 \\
  & Mostly-Uu & - & - & 1 & 3 \\
  \bottomrule
  \end{tabularx}
  \end{table}


\begin{figure}
    \centering
    \includegraphics[width=0.5\textwidth]{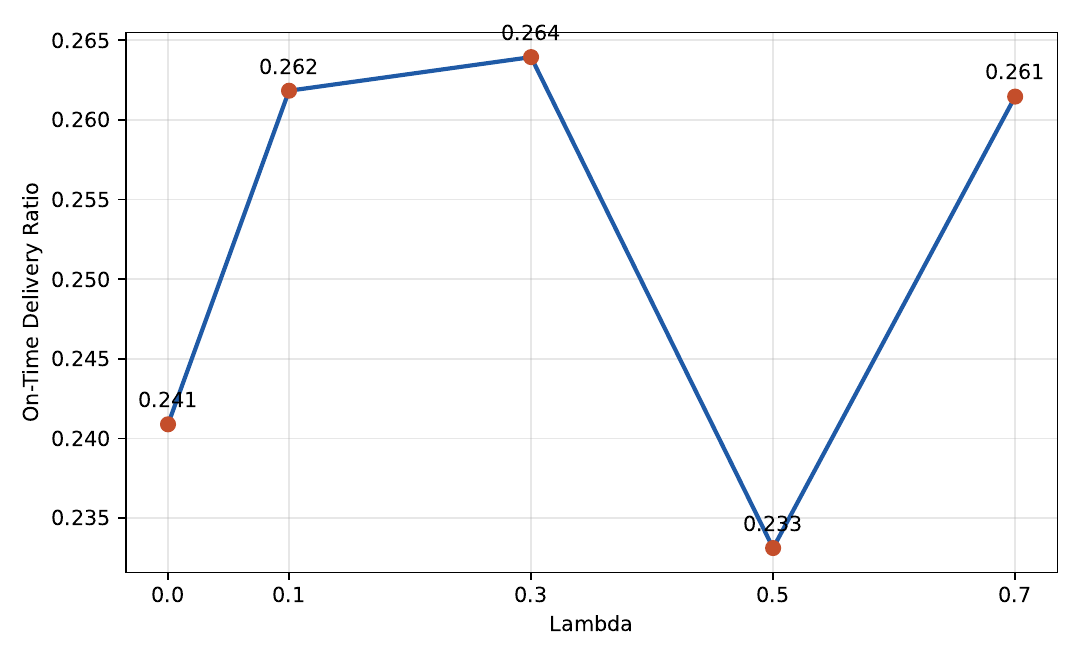}
    \caption{Impact of different $\lambda$ values on the mean on-time delivery ratio during training under the selected scenarios.}
    \label{fig:lambda}
\end{figure}

Figures \ref{fig:main-delivery} and \ref{fig:main-reliability} together with Table \ref{tab:main-results} show the overall results regarding the delivery and reliability of the models. The best checkpoint each of DDQN,MAPPO is used for illustrative purposes. 

To determine if MAPPO performs better than the DDQN model, we trained 10 models each with pairwise seeds and performed a two-sided paired t-test on the matched evaluation configurations. MAPPO achieved a higher mean delivery ratio in both controlled-vehicle settings.  Across both settings combined, MAPPO significantly outperformed DDQN (n = 320, MAPPO: M = 0.551, DDQN: M = 0.528, mean paired difference = 0.0227, t = 2.509, p = 0.0126). Here, n denotes the number of matched MAPPO--DDQN pairs, M the mean delivery ratio, t is the paired t-test statistic, and p is the corresponding two-sided p-value. The value n consists of 16 configurations × 2 controlled vehicle settings × 10 seeds = 320.
This means despite fewer environment steps, MAPPO converges faster.

In the comparison against the baseline methods we used the respectively best of the 10 trained models and compare it in  Figure \ref{fig:main-delivery}. MAPPO leads in the one-controlled vehicle setting. In the more relevant all-vehicles controlled setting the lead is even better visible.

In Figure \ref{fig:lambda} the impact of different lambda values on the on-time delivery ratio during training are displayed compared with different runs with the same seed. The best value was achieved at $\lambda = 0.3$, which we used for the model comparison.

An interesting finding is the occurrence of the wins by the static methods. The results are summarized in Table \ref{tab:method-win-counts}. The wins of the simpler strategies mostly occur in the CAM-only app-mix. Yacheur et al. \cite{10199400} observed that their MCDM RAT selection baseline approach achieves
better PRR in low-density traffic compared to LTE-V2X. Our observations confirm this finding, indicating that simple methods are most effective in basic scenarios. 

\section{Limitations}
We acknowledge several limitations of our study. 
Since we only used two maps, which are both from the same region we cannot determine the generalization property. Furthermore the DDQN model has some advantages, since it had longer training time due to the adaptive episode end and it has the hyperparameters from the Yacheur paper. Our MAPPO used default hyper-parameters for PPO provided by the RLlib library.
Additionally, the DDQN model inspired by Yacheur et al. and our MAPPO model use different reward functions. We did not conduct an ablation study regarding the reward function.

\section{Conclusion}
This paper presented a MARL based hybrid communication selection approach on heterogeneous QoS-dependent applications. We evaluate it against heuristics as well as DRL based previous work and found improvements in terms of on-time delivery ratio in our evaluation scenario.
In future work, we will apply the selection algorithm to a real-world testbed in order to examine the performance of the algorithm. Furthermore, in realistic scenarios a mix of adaptive and non-adaptive traffic should be analyzed.


\section*{Acknowledgment}
The authors would like to thank the German Federal Ministry of Research, Technology and Space (BMBFTR) which funded the project under grant number 16MEE0385 and the EU HORIZON-KDT Joint Undertaking, which funded the ShapeFuture project under the grant agreement n° 101139996.


\bibliographystyle{IEEEtran}
\bibliography{references}

@Article{s21030843,
AUTHOR = {Miao, Lili and Virtusio, John Jethro and Hua, Kai-Lung},
TITLE = {PC5-Based Cellular-V2X Evolution and Deployment},
JOURNAL = {Sensors},
VOLUME = {21},
YEAR = {2021},
NUMBER = {3},
ARTICLE-NUMBER = {843},
URL = {https://www.mdpi.com/1424-8220/21/3/843},
PubMedID = {33513998},
ISSN = {1424-8220},
DOI = {10.3390/s21030843}
}

@Article{fi16040107,
AUTHOR = {Khalid, Ihtisham and Maglogiannis, Vasilis and Naudts, Dries and Shahid, Adnan and Moerman, Ingrid},
TITLE = {Optimizing Hybrid V2X Communication: An Intelligent Technology Selection Algorithm Using 5G, C-V2X PC5 and DSRC},
JOURNAL = {Future Internet},
VOLUME = {16},
YEAR = {2024},
NUMBER = {4},
ARTICLE-NUMBER = {107},
URL = {https://www.mdpi.com/1999-5903/16/4/107},
ISSN = {1999-5903},
DOI = {10.3390/fi16040107}
}

@INPROCEEDINGS{10199400,
  author={Yacheur, Badreddine Yacine and Ahmed, Toufik and Mosbah, Mohamed},
  booktitle={2023 IEEE 97th Vehicular Technology Conference (VTC2023-Spring)}, 
  title={DRL-Based RAT Selection in a Hybrid Vehicular Communication Network}, 
  year={2023},
  volume={},
  number={},
  pages={1-5},
  doi={10.1109/VTC2023-Spring57618.2023.10199400}}

@article{RAVIGLIONE202470,
title = {ms-van3t: An integrated multi-stack framework for virtual validation of V2X communication and services},
journal = {Computer Communications},
volume = {217},
pages = {70-86},
year = {2024},
issn = {0140-3664},
doi = {https://doi.org/10.1016/j.comcom.2024.01.022},
url = {https://www.sciencedirect.com/science/article/pii/S0140366424000227},
author = {F. Raviglione and C.M. Risma Carletti and M. Malinverno and C. Casetti and C.F. Chiasserini}
}

@misc{pegurri2026van3twinmultitechnologyv2xdigital,
      title={VaN3Twin: the Multi-Technology V2X Digital Twin with Ray-Tracing in the Loop}, 
      author={Roberto Pegurri and Diego Gasco and Francesco Linsalata and Marco Rapelli and Eugenio Moro and Francesco Raviglione and Claudio Casetti},
      year={2026},
      eprint={2505.14184},
      archivePrefix={arXiv},
      primaryClass={cs.NI},
      url={https://arxiv.org/abs/2505.14184}, 
}

@inproceedings{dlr71460,
          editor = {Aida Omerovic, SINTEF \& University of Oslo and Diglio A. Simoni, RTI International - Research Triangle Park and Georgiy Bobashev, RTI International - Research Triangle Park},
            year = {2011},
       booktitle = {Proceedings of SIMUL 2011, The Third International Conference on Advances in System Simulation},
          author = {Behrisch, Michael and Bieker, Laura and Erdmann, Jakob and Krajzewicz, Daniel},
       publisher = {ThinkMind},
           title = {SUMO {--} Simulation of Urban MObility: An Overview},
           month = {Oktober},
             url = {https://elib.dlr.de/71460/},
            isbn = {978-1-61208-169-4}
}

@misc{ns3,
  title = {ns-3: Discrete-Event Network Simulator},
  howpublished = {\url{https://www.
nsnam.org/}},
  note = {Accessed: 2026-03-18}
}

@InProceedings{pmlr-v80-liang18b,
  title = 	 {{RL}lib: Abstractions for Distributed Reinforcement Learning},
  author =       {Liang, Eric and Liaw, Richard and Nishihara, Robert and Moritz, Philipp and Fox, Roy and Goldberg, Ken and Gonzalez, Joseph and Jordan, Michael and Stoica, Ion},
  booktitle = 	 {Proceedings of the 35th International Conference on Machine Learning},
  pages = 	 {3053--3062},
  year = 	 {2018},
  editor = 	 {Dy, Jennifer and Krause, Andreas},
  volume = 	 {80},
  series = 	 {Proceedings of Machine Learning Research},
  month = 	 {10--15 Jul},
  publisher =    {PMLR},
  url = 	 {https://proceedings.mlr.press/v80/liang18b.html}
}

@inproceedings{10.1145/3389400.3389404,
author = {Yin, Hao and Liu, Pengyu and Liu, Keshu and Cao, Liu and Zhang, Lytianyang and Gao, Yayu and Hei, Xiaojun},
title = {ns3-ai: Fostering Artificial Intelligence Algorithms for Networking Research},
year = {2020},
isbn = {9781450375375},
publisher = {Association for Computing Machinery},
address = {New York, NY, USA},
url = {https://doi.org/10.1145/3389400.3389404},
doi = {10.1145/3389400.3389404},
booktitle = {Proceedings of the 2020 Workshop on Ns-3},
pages = {57–64},
numpages = {8},
location = {Gaithersburg, MD, USA},
series = {WNS3 '20}
}

@techreport{gaa2025cv2xvol1,
  title       = {C-V2X Use Cases and Service Level Requirements, Volume I},
  author      = {{5GAA Automotive Association}},
  institution = {5G Automotive Association (5GAA)},
  year        = {2025},
  month       = jan,
  number      = {Version 2.0},
  day         = {14}
}

@inproceedings{10.5555/3600270.3602057,
author = {Yu, Chao and Velu, Akash and Vinitsky, Eugene and Gao, Jiaxuan and Wang, Yu and Bayen, Alexandre and Wu, Yi},
title = {The surprising effectiveness of PPO in cooperative multi-agent games},
year = {2022},
isbn = {9781713871088},
publisher = {Curran Associates Inc.},
address = {Red Hook, NY, USA},
booktitle = {Proceedings of the 36th International Conference on Neural Information Processing Systems},
articleno = {1787},
numpages = {14},
location = {New Orleans, LA, USA},
series = {NIPS '22}
}

@INPROCEEDINGS{10881343,
  author={Mancini, Lorenzo and Labbi, Safwan and Meraim, Karim Abed and Boukhalfa, Fouzi and Durmus, Alain and Mangold, Paul and Moulines, Eric},
  booktitle={2024 IEEE Middle East Conference on Communications and Networking (MECOM)}, 
  title={Joint Channel Selection using FedDRL in V2X}, 
  year={2024},
  volume={},
  number={},
  pages={268-273},
  doi={10.1109/MECOM61498.2024.10881343}}

@INPROCEEDINGS{9868865,
  author={Chen, Weixiang and Gu, Bo and Tan, Xiaojun and Wei, Chenhua},
  booktitle={2022 International Conference on Computer Communications and Networks (ICCCN)}, 
  title={Radio Resource Selection in C-V2X Mode 4: A Multiagent Deep Reinforcement Learning Approach}, 
  year={2022},
  volume={},
  number={},
  pages={1-8},
  doi={10.1109/ICCCN54977.2022.9868865}}
\end{document}